\documentclass[9pt]{article}
\usepackage{spconf,amsmath,graphicx}
\usepackage{hyperref}
\usepackage{enumitem}

\title{DLR: Toward A Deep learned rhythmic representation\\for music content analysis}

\name{Yeonwoo Jeong\textsuperscript{$\dag$}, Keunwoo Choi\textsuperscript{*}, Hosan Jeong\textsuperscript{$\dag$}}
\address{\textsuperscript{$\dag$}EECS, Seoul National University,
Seoul, Republic of Korea, \\
\textsuperscript{*}Centre for Digital Music, EECS, Queen Mary University of London, London, UK\footnotemark[2]\\
\texttt{legend4020@snu.ac.kr}}
\ninept 
\begin{document}
\maketitle
\begin{abstract}
In the use of deep neural networks, it is crucial to provide appropriate input representations for the network to learn from. In this paper, we propose an approach to learn a representation that focus on rhythmic representation which is named as \textit{DLR} (Deep Learning Rhythmic representation). The proposed approach aims to learn DLR from the raw audio signal and use it for other music informatics tasks. A 1-dimensional convolutional network is utilised in the learning of DLR. In the experiment, we present the results from the source task and the target task as well as visualisations of DLRs. The results reveals that DLR provides compact rhythmic information which can be used on multi-tagging task.

\end{abstract}
\begin{keywords} 
deep neural networks, music, rhythm
\end{keywords}

\section{Introduction}
\label{sec:intro}
Deep learning approaches are often called `end-to-end', which means the learning procedure optimises the relevant representations. Nevertheless, providing appropriate input representation is one of the important design choices which involves two contrary goals; First, \textit{efficiency} -- the representations should be compact rather than redundant so that the training focuses on the learning that can outperform hand-written features; Second, \textit{effectiveness} -- the representation should not be the information bottleneck but allow the network to extract the relevant information from.

In music information retrieval (MIR) research, majority of researches has been relying on 2-dimensional time-frequency representations \cite{choi2017tutorial}. They decompose an audio signal by various frequencies and their properties are advantageous to provide spectral or harmonic information. For example, Mel-spectrograms provide an efficient representation by compressing the frequency axis (which is usually of 256/512/1024 bins) into Mel-frequencies (usually up to 128 Mel-bins). Constant-Q transform is preferred for its log-equidistant centre frequencies which enables an intuitive harmonic-invariant learning.

Rhythmic information have shown to be useful in many MIR tasks, e.g., music genre classification \cite{tzanetakis2002musical} and measuring music similarity \cite{pohle2009rhythm} and it has lead to deep learning approaches that aims to focus on rhythmic aspects of music. For example, 'wide' kernels of convolution layers
that span over long time duration are used in \cite{pons2017designing} and temporal patterns are aggregated with fully-connected layers in \cite{jeong2016learning} in order to learn rhythmic patterns, but both of them are based on 2-dimensional representations (Mel-spectrogram and cepstrogram).

A further data-driven approach is to \textit{learn} a new musical representation that focus on rhythmic aspects, probably from the original 1-dimensional time-series representation of an audio signal (often referred to \textit{raw audio}). There have been researches on learning a new representation, although none of them explicitly focus on the rhythmic aspect to our best knowledge. On learning from raw audio, \cite{dieleman2014end} introduced a frame-based feature learning that resulted in a Mel-spectrogram-like conversion. More recently, \cite{lee2017sample} proposed a sample-based learning, achieving state-of-the-art performance in music tagging. There are also related approaches that learn features from a 2-dimensional time-frequency representation. \cite{korzeniowski2016feature} proposed to learn a chromagram, which is rather a task-specific music feature while \cite{choi2017transfer} introduced a general-purpose one.

In this paper, we introduce to learn a representation that focus on the rhythmic aspect of music signal. The proposed representation can be \textit{transferred} and used as a input representation for MIR tasks. Section \ref{sec:bg} provides the backgrounds on dilated convolution, a special type of convolution operations, as well as the existing rhythmic representations. The proposed method, deep learned rhythmic representation (DLR), is introduced in Section \ref{sec:proposed}. We present and discuss the experiment results in Section \ref{sec:exp} and conclude our work in Section \ref{sec:conc}.


\section{Backgrounds} \label{sec:bg}

A \textit{dilated convolution} is a special type of convolution and specified by the dilation rate $d$ and illustrated in Figure \ref{fig:source_task}. It is identical to normal convolutional layer When $d$=1. With an integer $d>1$, the kernel skips $d-1$ samples on the input side to learn patterns in a wider range while keeping the same number of the weights (3 in the Figure \ref{fig:source_task}). 
It has been used to learn directly from raw audio input \cite{oord2016wavenet}. 

\textit{Tempogram} is a mid-level 2-dimensional rhythmic representation that encodes local tempo information by computing local auto-correlations of estimated onset strengths \cite{grosche2010cyclic} by Grosche et al. The auto-correlation with a lag of $\tau$ affects to that with a lag of $n \tau$ and vice versa, and Grosche et al. proposed to focus on the tempi differing by a power of two in order to reduce this effect. 

\textit{SuperFlux} \cite{bock2013maximum} was proposed to estimate the onsets in music signal by B\"ock and Widmer and is based on modifications of Spectral flux and few signal processing techniques. A spectral flux is related to onsets and originally computed by $\sum_k H(|X(n, k)|-|X(n-\mu, k)|)$ with $\mu=1$ where $X$ is a spectrogram, $H(x)=\frac{x+|x|}{2}$, and $n$, $k$ refer to the time, frequency indices respectively. With an increased temporal resolution of the magnitude spectrogram, B\"ock and Widmer proposed to compute a spectral flux with $\mu>1$. This can be viewed as as a \textit{dilated} spectral flux. Later, an \textit{accent signal}, which is a variant of SuperFlux, was used in \cite{prockup2015modeling}, combined with Mellin scale transform to obtain a tempo-invariant rhythmic representation.

\section{The proposed approach} \label{sec:proposed}

\begin{figure}[t]
	\centering   \includegraphics[width=\columnwidth]{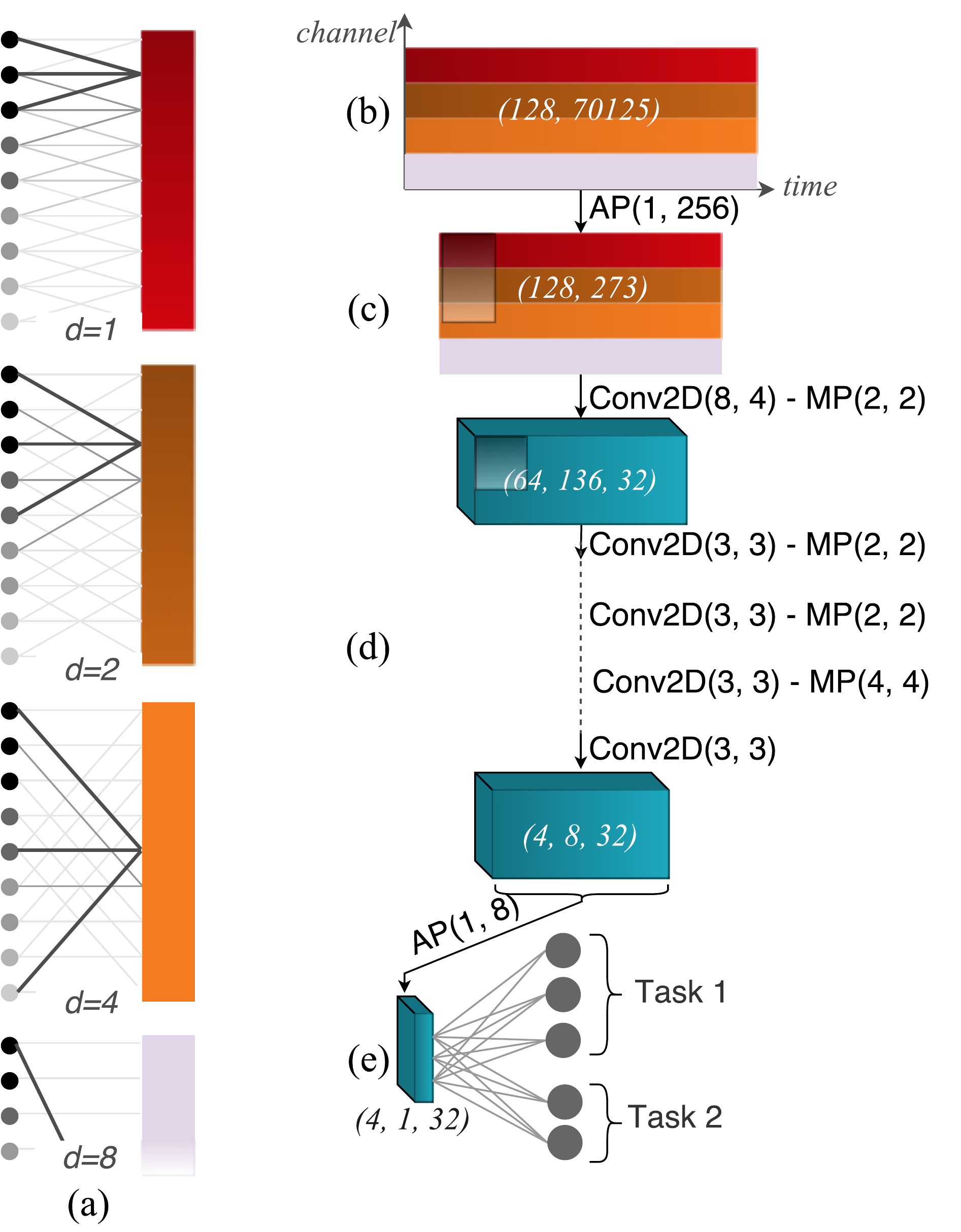}
	\caption{A block diagram of the procedure to train DLR. (a) A 1D dilated convolutions with various dilation rates $d$ are performed in parallel, then (b) concatenating the feature maps, and (c) average pooling is applied along time axis. This feature maps will be used as a DLR after training. Note that the procedure after (c) is subject to change depending on the used source task. (d) In our experiment, a 5-layer 2D convnet is used, then (e) a global average pooling is applied along time axis, followed by dense layers for two tasks - genre and tempo classifications.}
	\label{fig:source_task} 
\end{figure}

Based on the background survey, we suggest several desired aspects of the learned rhythmic representation as follows:
\begin{itemize}[leftmargin=*]
\item Temporality : representing the \textit{whenabouts} of rhythmic aspects
\item Precise temporal resolution: more than usual time-frequency representation to be useful and not redundant
\item Multiple temporal resolutions: since rhythmic events are often repeatitive, if not always, at different rates.
\end{itemize}

These aspects lead us to propose a framework to obtain such a representation which is named as \textit{DLR} (Deep Learned Rhythmic representation). 
DLR can be learned by taking advantage of a rhythm-oriented source task as a proxy task. The learning procedure is illustrated in Figure \ref{fig:source_task}, of which details are as below.

\begin{enumerate}[label=(\alph*), leftmargin=0.6cm]
  \setlength\itemsep{0em}
  \item On the raw audio input, convolution layers with three dilation rates are used, yielding in three feature maps with focussing on local and temporal relationships with different resolution. The power-of-2 dilation is similar to the structure of WaveNet \cite{oord2016wavenet} and the computation process of the tempogram. We use the power-of-$\alpha$ dilation in this experiment. We assume it provides higher-level representation then a tempogram, although it is hard to directly compare them. Unlike \cite{prockup2015modeling}, DLR is not tempo-invariant which may be implemented in using it. 
  
  \item The feature maps are concatenated along channel-axis. In the figure, we assumed the time length to be 70,125 as an example.
  
  \item The concatenated feature maps are subsampled along time-axis using \texttt{average} operation (average pooling) with the rate and the stride of 256. As a result, it becomes the DLR, with a size of (128,~273) as an example.
  
  \item After the feature map in Figure \ref{fig:source_task} (c) which will serve the role as DLR (after training), any network structure can follow if it provides a backpropagation with respect to rhythmic information of the input.\\In our experiment, we use 2-dimensional convolution networks (convnets) proposed in \cite{choi2016automatic} . Although the channel axis of DLR is rather in an arbitrary order than with some spatiality, e.g., frequency axis, 2-dimensional convnet achieved better performance than 1-dimensional one \cite{dieleman2014end}, possibly because the distortion invariance along the both axes resulted in more flexible learning. 

\item In our implementation, a global average pooling is applied to aggregate the information over time, after which two dense layers follow as output layers since we use a multi-task problem as will be explained in Section \ref{subsec:source_task}.
\end{enumerate}


One of the hypotheses of preliminary experiments was whether the amplitude of the audio signal is more effectively represented in the decibel scale, i.e., $\log_{10}( \max(|x|, \epsilon))$. The results showed that decibel scaling slightly degrades the accuracy on our source task, therefore we propose to use the original amplitude as it is.

\section{Experiments and discussions} \label{sec:exp}

This section presents and discusses the details of two experiments: \textit{i)} training DLRs (source task) and \textit{ii)} applying it to the music tagging (target task). Since the goal is to learn a new representation that is applicable for various target music tasks, we do not prioritize to obtain the best performance on the source task. Instead, we control the size of the DLR regarding its efficiency in the target tasks. Limiting the size is also related to a correct comparison between DLR and other representations. This is because the input representation size may affect the network structure (which may lead to affect the representation power of the network).
In the target task, in detail, we measure how much information DLR provides \textit{i)} by itself and \textit{ii)} along with a Mel-spectrogram and \textit{iii)} how advantageous it is compared to tempogram.

For both tasks, the audio signal is downmixed to mono and downsampled to 8~kHz. The sampling rate is set to be relatively low for an efficient training, assuming the primary rhythmic events exist under 4~kHz. 
We use Tensorflow \cite{abadi2016tensorflow} as a deep learning framework. Scikit-learn \cite{pedregosa2011scikit} and Librosa \cite{mcfee2017librosa} is used for machine learning procedure such as splitting and audio signal processing respectively. The proposed representation is released online. \footnote{\url{http://github.com/keunwoochoi/DLR}}


\subsection{Source Task} \label{subsec:source_task}
As mentioned in Section~\ref{sec:proposed}, training DLR requires a source task that focuses on the rhythmic aspect of music. In this paper, we use Extended ballroom dataset \cite{marchand2016extended} which is provided with 4,180 audio items and \textit{i)} their genre labels which primarily depend on their rhythmic properties and \textit{ii)} tempo annotation in BPM (beat per minute). The dataset includes four minor genres with relatively small number of samples (53, 47, 65, 23 items) which are excluded in our experiments for more balanced training. As a result, 3,992 items and 9 genres are used, which are split into 3,232/360/400 items for training/validation/test set respectively. We use the genre labels to perform the stratified random split.

In order to exploit the dataset, we configure a multi-task problem consisting of genre and tempo classification. In the genre classification, only the nine major genres are used due to the lack of the training data for the four minor genres. With the tempo label, we modify the problem as a \textit{classification} rather than a regression problem since it is more straightforward for evaluation which is crucial to obtain a meaningful DLR. After clustering the tempi of the dataset, we categorise the data sample into four tempo classes with boundaries of 112, 149, 187 BPM. Cross-entropy is used as a loss function for both genre and tempo classification tasks. During training, the learning rate is adaptively controlled by ADAM optimizer \cite{kingma2014adam}. 

Regarding selecting the parameters of the dilated convolution, one of the constraints is to set the size of DLR to be comparable to other representation such as Mel-spectrogram. The output size of DLR is fixed in channel direction, the filter length of 1-dimensional convolution network is fixed, and the convolution network used to perform source task is fixed to compare the performance of DLR.

We define $d$, the dilation parameters that governs the whole dilated convnet. $d$ are specified by a set of the specific dilation scales, $\alpha$'s. For example, \texttt{DLR(5)} means $\alpha={1, 5, 25, 125}$ for the four convolutional layers.

To investigate the performance by varying dilation rates $d$, we configure three different DLRs: $d=1, 5, 13$. We presume that varying dilation rate leads to representations that focus on the temporal patterns with different rate of changing over time.  \texttt{DLR(1)} is set to 1 as a baseline method which means a dilation rate of a 1-dimensional convolutional network is 1. The number of convolutional layers ($n$) 
is set to 4. 
$\alpha$ is set to 5 and 13 in \texttt{DLR(5)} and \texttt{DLR(13)} respectively. 
These result in the corresponding dilation rates ($d$) of each layer to be (1, 5, 25, 125) and (1, 13, 169, 2,197) to confirm which dilation scale ($\alpha$) is efficient for DLRs to contain diverse temporal patterns. 

\subsubsection{Results and Discussion} 
Table \ref{table:dlr_accs} summarises the performances on the source task with different DLR parameters. 
As shown in Table \ref{table:dlr_accs}, all three configurations showed competitive performances. \texttt{DLR(13)}, the one with $\alpha=13$ achieved the best performance in both genre and tempo classification. 

DLRs with a small $\alpha$ might focus on trivial, a very short-term temporal patterns, which might be ineffective. On the other hand, DLRs with a large $\alpha$ might end up ignoring some precise temporal patterns. We found the setup with $\alpha=13$ is sensible, where the time resolution of convolutional networks rate is approximately 0.27 second $(=13^3/8000)$ which is sufficiently large to capture the changes of musical events.

As a sanity check of the source task structure, we present the genre classification and tempo classification accuracies. \cite{choi2017transfer} reports 0.867 of genre classification accuracy, and \cite{marchand2016scale} reports 0.949 of genre classification recall while our structure achieves 0.927 to 0.937 accuracy on major 9 genres in the Extended ballroom dataset. Also, \cite{schreiberpost} reports 0.904 as tempo estimation accuracy which allows 4\% of tolerance while our structure achieves 0.952 to 0.970 as a accuracy on tempo classification with 4 classes.

\begin{table}
\begin{center}
\caption{The accuracies of different DLRs on the source task. The triples of the parameters indicate (the number of total convolutional layers($n$), their feature map numbers(n\_channel), dilation scale($\alpha$) between dilation rates $d \in \{1, \alpha, \cdots, \alpha^{n-1}\}$). \textit{i)} The number of total convolutional layers times their feature map numbers to be total feature map numbers(128), \textit{ii)} The dilation scale adjusts the gap between dilation rates ($d$), \textit{iii)} Filter length of each convolutional networks is fixed with 16.
} \label{table:dlr_accs}

\begin{tabular}{@{}c|c|c|c|@{}}
index & \begin{tabular}[c]{@{}l@{}}DLR parameters\\ ($n$, n\_channel, $\alpha$)\end{tabular}& Genre & Tempo \\\hline \hline
\texttt{DLR(1)} & (1,128, 1) &  $0.932$ & $0.952$ \\ \hline 
\texttt{DLR(5)} & (4, 32, 5) &  $0.927$ & $0.955$  \\ \hline
\texttt{DLR(13)}& (4, 32,13) & $0.937$ & $0.970$  \\ 
\end{tabular}
\end{center}
\end{table}

\subsection{Target Task}

\begin{figure}[t]
	\centering   \includegraphics[width=0.85\columnwidth]{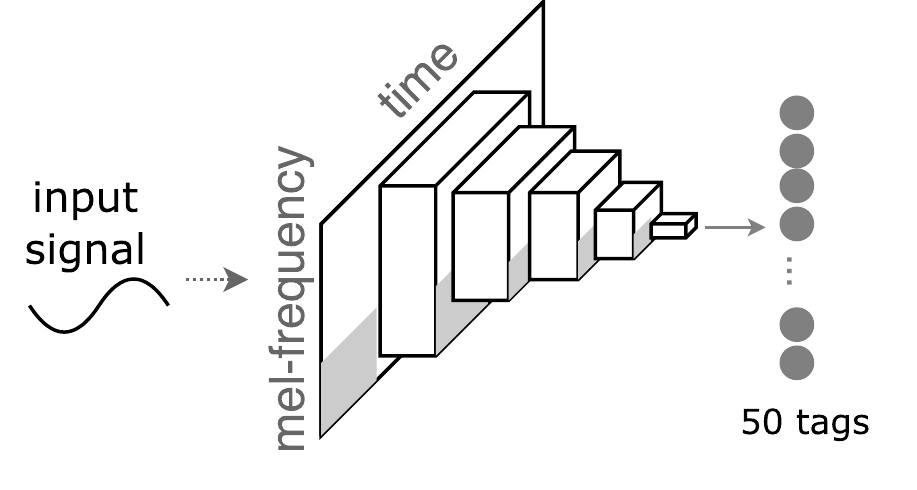}
	\caption{Block diagram of the 2-dimensional 5-layer convnet that is used for a Mel-spectrogram input. Convolutional layers use kernel sizes of $4\times 8$ in the first layer and $3\times 3$ for the 2nd-5th layers. Each convolutional layer is followed by Batch normalization, ReLU, and max-pooling layers.}
	\label{fig:target_task} 
\end{figure}

We select music tagging on Million song dataset \cite{bertin2011million} as a target task. Music tagging is a multi-label classification problem where an item can be positively tagged by multiple labels. The tags include genres, instruments, moods, and eras \cite{choi2017effects}. Many rhythmic aspects are related to those tags, therefore we assume that the performance on music tagging can be used as a measure the used input representation. 244,222 items that are tagged with at least one positive label in top-50 popular tags are used and split into training/validation/test set (200,860/12,540/28,390), following the configuration in \cite{choi2016automatic} and the provided split sets\footnote{\url{https://github.com/keunwoochoi/MSD_split_for_tagging}}. 
In the signal, the first 29-second is converted into Mel-spectrograms with 128 Mel-bins, resulting in the Mel-frequency resolution of about 18.5 Hz under 1 kHz. The magnitude is mapped by decibel scale with a dynamic range of 80 dB, limited by the maximum decible value, following the best configuration in \cite{choi2017comparison}.


A 2-dimensional convnet is used on the Mel-spectrogram inputs. The structure is similar to one in \cite{choi2016automatic}, i.e., a fully-convolutional network with square-shaped convolutional kernels. Considering the increased frequency resolution, we use $4\times 8$ kernels on the first layer, and $3\times 3$ ones on the other layer. All convolutional layers have 32 channels and are followed Batch normalization \cite{ioffe2015batch}, ReLU as activation function, and max-poolings of [(2, 2), (2, 2), (2, 2), (4, 4)].

\subsubsection{Results and Discussion} 

\begin{figure}[t]
\centering
  \includegraphics[width=1.0\columnwidth]{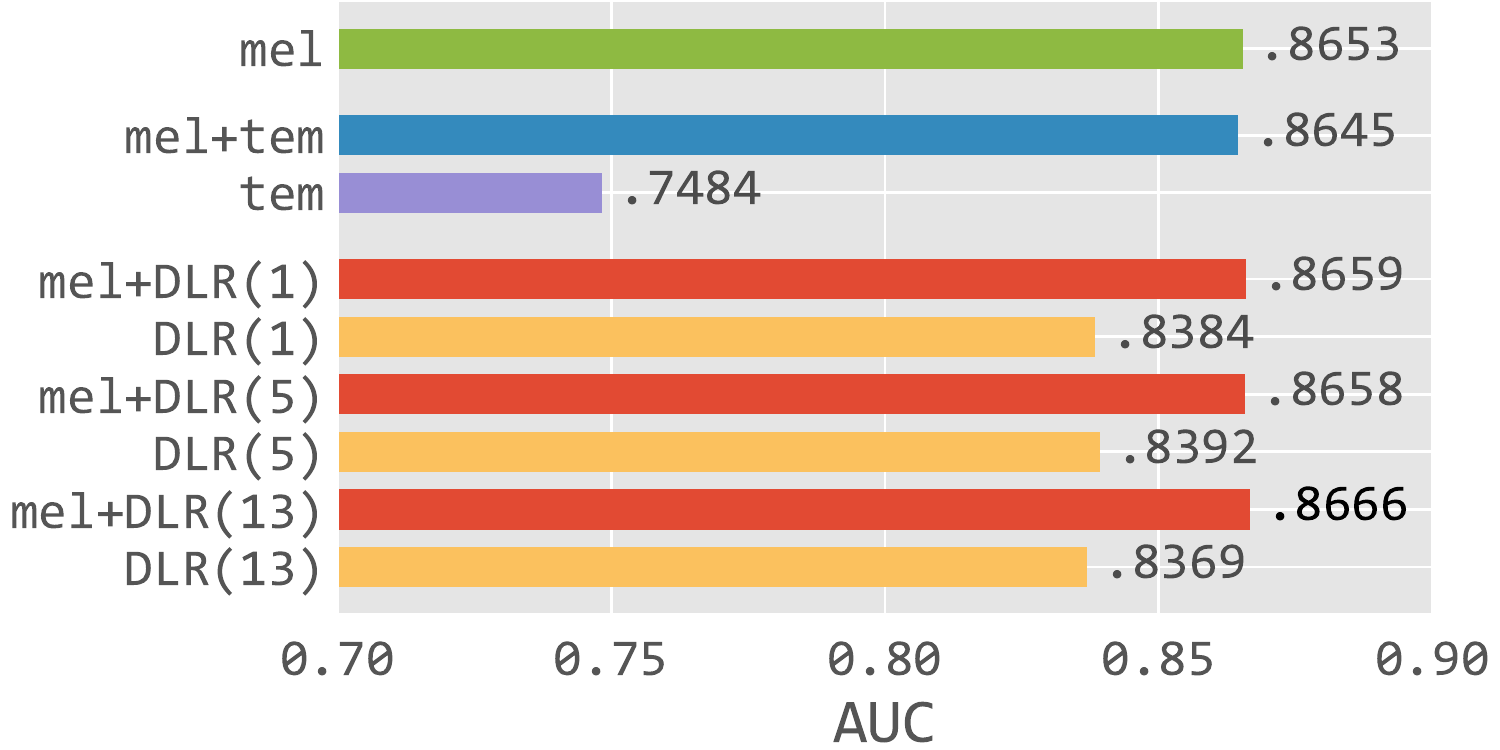}\\ %
	\caption{Target task performances in AUC with different input representations and their combinations. \texttt{mel} and \texttt{tem} corresponds to Mel-spectrogram and tempogram respectively.}
	\label{fig:target_result}
\end{figure}

\begin{figure}[t]
	\centering           \includegraphics[width=\columnwidth]{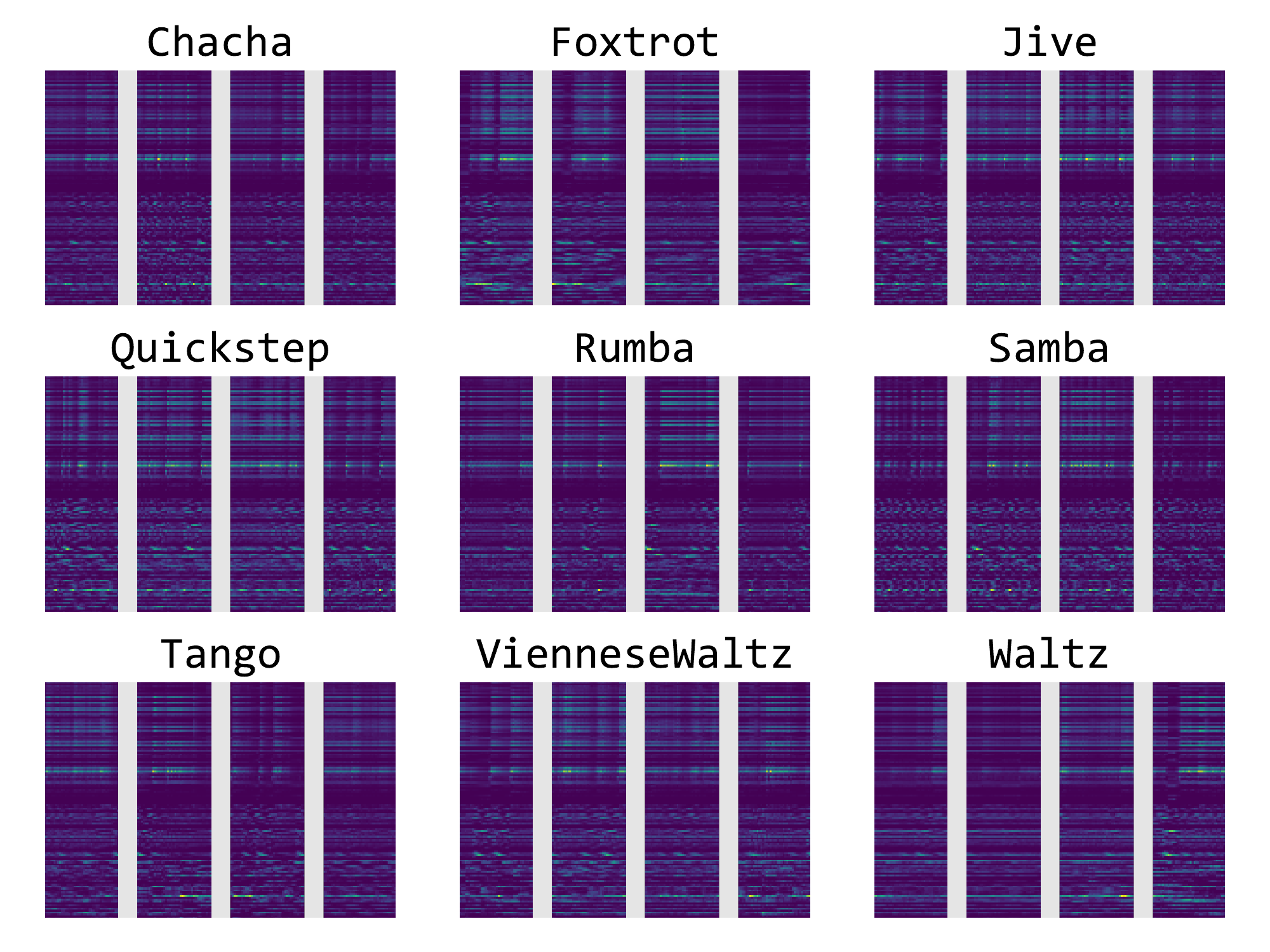}
	\caption{Visualisations of \texttt{DLR(13)} on nine major genres in Extended ballroom dataset -- DLRs of four selected items in each genre. Each DLR represent a short segment with 40 frames (which is up to 65 ms) and concatenating the feature maps with different dilation rate, $d$, i.e., a trimmed part of (c) in Figure \ref{fig:source_task}.}
	\label{fig:visualisation} 
\end{figure}


Experiment was set to compare tempogram \cite{grosche2010cyclic}  and DLR that is transferred from the source task. Figure \ref{fig:target_result} shows their single performances and performances with Mel-spectrogram on target task. All of DLRs shows performance comparable to Mel-spectrogram with small margin (.0161 to .0196) while the network with tempogram shows relatively low performance with the margin (.1169). Compared to the network with tempograms, ones with DLRs achieved significantly better performance (.0873 to .0908) even though the size of DLR is only half the size of tempogram, implying a DLR efficiently provides richer information than a tempogram does.


When used along with a Mel-spectrogram, our implementation of DLRs did not seem to provide complimentary information, which is probably similar with tempogram. As shown in Table\ref{table:dlr_accs} and Figure \ref{fig:target_result}, adding DLRs to Mel-spectrogram slightly improved the performance by 0.0005 to 0.0016. %


The Figure \ref{fig:visualisation} shows the \texttt{DLR(13)} for 9 genres in Extended ballroom dataset. For each genre, 4 DLRs are selected which are top-4 mostly classified as each corresponding genre among the Extended ballroom dataset by the network in the source task. The top and the bottom part of the DLRs come from small and large dilate rate respectively.

In each DLR, the upper part of DLRs generally shows similar pattern along nine genres. We presume this due to small dilation rate resulted in learning very similar pattern in each channel. On the other hand, the large dilation part (the bottom of DLRs) is rather different/diverse, which may mean longer-term patterns can be more useful for representing relevant musical aspect.

\section{Conclusions} \label{sec:conc}
In this paper, we introduced \textit{DLR}, a deep rhythmic representation that is learned and transferable to other tasks. We suggested three aspects of the desired rhythmic representation, based on which we designed our DLR. 
In the experiment, DLR was first trained for a rhythm- and tempo-related task and transferred as an input representation to a convolutional neural network for music tagging. The results showed that DLR is more efficient and compact representation on multi tagging task than tempogram and the network with DLR was nearly as strong as the one with Mel-spectogram. However, when used together with Mel-spectogram, DLR showed a slight improvement on performance. The future work will further investigate to steer DLR to be more complimentary to existing representations and analyse the properties of the learned representation.
\bibliographystyle{IEEEbib}
\bibliography{ref}

\end{document}